\documentclass[prl, twocolumn,superscriptaddress,nobibnotes]{revtex4-2}
\usepackage{graphicx}
\usepackage{amssymb}
\usepackage{amsmath}
\usepackage{bm}
\usepackage{color}
\usepackage{float}
\usepackage{setspace}
\usepackage{physics}
\usepackage{subfigure}

\large

\begin{document}
	%\linenumbers
	\title{Observation of condensed moir\'e exciton polaritons \\in twisted photonic lattices at room temperature}
	
	\author{Chunzi Xing}
	\affiliation{Department of Physics, School of Science, Tianjin University, Tianjin 300072, China}
	
	\author{Yu Wang}
	\affiliation{The MOE Key Laboratory of Weak-Light Nonlinear Photonics and International Sino-Slovenian Join Research Center on Liquid Crystal Photonics, TEDA Institute of Applied Physics and School of Physics, Nankai University, Tianjin 300457, China}
	
	\author{Tobias Schneider}
	\affiliation{Department of Physics and Center for Optoelectronics and Photonics Paderborn (CeOPP), Universit\"{a}t Paderborn, 33098 Paderborn, Germany}
	
	\author{Xiaokun Zhai}
	\affiliation{Department of Physics, School of Science, Tianjin University, Tianjin 300072, China} 
	
	\author{Xinzheng Zhang}
	\affiliation{The MOE Key Laboratory of Weak-Light Nonlinear Photonics and International Sino-Slovenian Join Research Center on Liquid Crystal Photonics, TEDA Institute of Applied Physics and School of Physics, Nankai University, Tianjin 300457, China}
	
	\author{Haitao Dai}
	\affiliation{Department of Physics, School of Science, Tianjin University, Tianjin 300072, China}
	
	\author{Xiao Wang}
	\affiliation{College of Materials Science and Engineering, Hunan University, Changsha 410082, China}
	
	\author{Anlian Pan}
	\affiliation{College of Materials Science and Engineering, Hunan University, Changsha 410082, China}
	
	\author{Zhenyu Xiong}
	\affiliation{Lab of quantum detection and awareness, Space Engineering University Beijing 101416, China}
	
	\author{Hao Wu}
	\affiliation{Lab of quantum detection and awareness, Space Engineering University Beijing 101416, China} 
	
	\author{Yuan Ren}
	\affiliation{Lab of quantum detection and awareness, Space Engineering University Beijing 101416, China} 
	
	\author{Stefan Schumacher}
	\affiliation{Department of Physics and Center for Optoelectronics and Photonics Paderborn (CeOPP), Universit\"{a}t Paderborn, 33098 Paderborn, Germany}
	\affiliation{Institute for Photonic Quantum Systems (PhoQS),
		Paderborn University, 33098 Paderborn, Germany}
	\affiliation{Wyant College of Optical Sciences, University of Arizona, Tucson, AZ 85721, USA}
	
	\author{Xuekai Ma}
	\affiliation{Department of Physics and Center for Optoelectronics and Photonics Paderborn (CeOPP), Universit\"{a}t Paderborn, 33098 Paderborn, Germany}
	
	\author{Tingge Gao}
	\affiliation{Department of Physics, School of Science, Tianjin University, Tianjin 300072, China}

	\begin{abstract}
		Moir\'e lattices attract significant attention in double-layer graphene and TMD layer heterostructures as well as in photonic crystals due to the interesting exotic physics that emerges within these structures. However, direct measurement of the moir\'e ground, excited states and Bloch bands in twisted photonic lattices is still illusive. In this work we report strong coupling between excitons in CsPbBr$_3$ microplates and moir\'e photonic modes at room temperature. Depending on the coupling strength between the nearest potential sites, we observe staggered moir\'e polariton ground states, excited states and moir\'e polariton bands. Phase locked moir\'e zero (in-phase) states and moir\'e $\pi$ (antiphase) states with different spatial distributions are measured. The moir\'e polariton distribution can be tuned into the shape of a parallelogram by controlling the depth and width of the potential in one photonic lattice with another superimposed one fixed. In addition, 
		moir\'e polaritons in twisted 2D honeycomb lattices are also observed. Increasing the pumping density, we realize exciton polariton condensation in the moir\'e potential sites of the 1D/2D twisted lattices with the coherence time of around 1.4 ps. Our work lays the foundation to study coherent moir\'e polariton condensation in twisted photonic lattices at room temperature.
		
		%trapped in the potential sites and across the twisted photonic lattices

	\end{abstract}
	
	%With different exciton and moir\'e photonic components, polaritons distribution can be tuned among the twisted photonic lattices. distribution in the shape of isolated dots, dimers and delocalized lines are formed.
	
	\maketitle
	
	%e observe the phase locking of zero (stable in-phase state) and $\pi$ (metastable antiphase state) between the moir\'e potentials with different moir\'e patterns. Remarkably, we observe the phase locking of moir\'e zero (stable in-phase) states and moir\'e $\pi$ (metastable antiphase) states with different spatial distributions across the twisted photonic lattices. By modifying the effective refractive index distribution of one photonic lattice, randomly distributed isolated exciton polariton dots, dimers and lines are observed.
	
	%\section*{Introduction}
	
	Moir\'e lattices in  graphene bilayer and TMD monolayer heterostructures offer a tunable platform where the interlayer coupling of electrons and excitons leads to a plethora of physical phenomena, for example, the transition from superconductor to Mott insulator \cite{caoyang}, observation of the quantum Hall effect \cite{moire quantum hall effect} and nontrivial gauge field \cite{moire gauge potential}, the appearance of moir\'e excitons \cite{exiton moire1, exiton moire2, exiton moire3}, and emergence of correlated exciton insulators \cite{flat band bilayer1, flat band bilayer2, exciton flat band}. On the other hand, the bands resulting from the coupling between potential sites arranged on a moir\'e lattice have been extended to photonic systems. In these systems, the localization and delocalization of light and nonlinear optical solitons can be observed by adjusting one photonic lattice against a second one into a particular angle \cite{photonic moire lattice}. The period of the moir\'e lattice in these twisted photonic structures can reach the size of several micros, thus the bands due to the coupling between the moir\'e potentials can be probed within several $\mu$m$^{-1}$ in angle-resolved spectroscopy. In this case, both the momentum space and real space images can be directly measured. For example, photonic bands in the optical frequency range are probed in a double photonic crystals \cite{moire band}. Cold atom condensates loaded in twisted photonic lattices have also been realized and the transition from superfluid to Mott insulator in the created moir\'e lattice is shown \cite{atom moire lattice}.

	However, direct measurement of the moir\'e ground state, excited states and moir\'e band states remain unexplored in the normal twisted photonic lattices. The observation of these modes can be used to investigate distinct state distribution in the momentum and real space resulted from different coupling strength among the moir\'e potentials. In addition, phase locking between different potential sites can be observed in the moir\'e bands. For example, the phase synchronization within a moir\'e flat band can enhance the performance of nanocavity laser array greatly \cite{flat band phase}. On the other hand, $\pi$ phase locking at the top of the moir\'e ground bands and bottom of the moir\'e excited bands will lead to different mode distribution compared to the ground band, which offers to study quantum simulation with tunable phase modulation. These $\pi$ phase locked moir\'e states have not been observed yet.

	\begin{figure}
		\centering
		\includegraphics[width=\linewidth]{./Fig1.pdf}
		\caption{\textbf{Schematics of the microcavity and photoluminescence exciton polariton dispersions in the moir\'e lattice.}(a) The structure of the microcavity. (b) The optical image of the twisted photonic structures, the right panel is the zoomed schematic of the moir\'e pattern. Photoluminescence dispersion along (c) $k_x$ and (d) $k_y$ direction. The dash lines indicate the edge of the first Brillouin zones. Moir\'e ground state (I), excited state (II) and moir\'e ground bands (III) are labeled in (c). }
	\end{figure}
	
	\begin{figure}
		\centering
		\includegraphics[width=\linewidth]{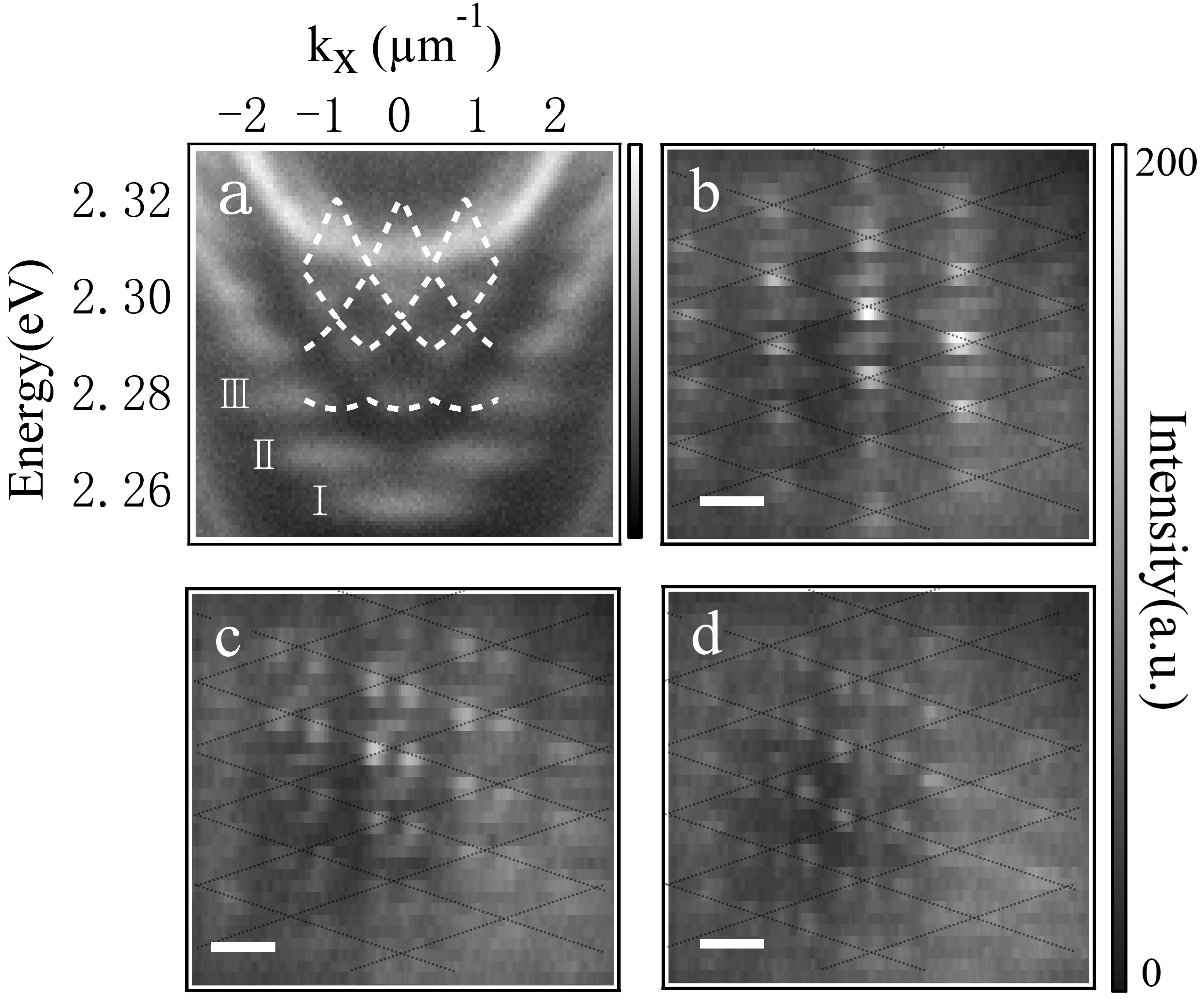}
		\caption{\textbf{Moir\'e exciton polariton patterns with different energies.} (a) Zoomed dispersions along $k_x$ with simulated dashed lines. Moir\'e ground state (I), excited state (II) and moir\'e ground bands (III) are labeled. (b-d) Moir\'e polariton spatial distributions of the ground state (I), the first excited state (II) and ground band states (III) labeled in Figure 2(a). The dash lines represent two photonic lattices. Scale bars in (b-d): 5 $\mu$m. }
	\end{figure}

	As a quasiparticle due to the strong coupling between the exciton and photon mode \cite{microcavity book} in Fabry-Perot microcavity, exciton polaritons can demonstrate similar Bose-Einstein condensation process as the cold bosonic particles at much higher temperature due to the very light effective mass and large exciton binding energy \cite{polariton BEC1, polariton BEC2, polariton GaN}. More importantly, exciton polaritons allow to investigate different phase locking when they are loaded into a photonic lattice \cite{yamamoto}. To observe phase locking of zero and $\pi$ of exciton polaritons in the moir\'e lattice at room temperature, specific semiconductor materials with robust excitons and easy integration with photonic structures are needed, among which the perovskite CsPbBr$_3$ microplates stand out with large exciton binding energy and oscillator strength \cite{Yao RD, Xiaokun vortex, gy pvk, lhh pvk, polariton pvk1, polariton pvk2}. 
	
	%Notably room temperature exciton polariton condensation based on CsPbBr$_3$ microplates has been demonstrated \cite{Yao RD, Xiaokun vortex, gy pvk, lhh pvk, polariton pvk1, polariton pvk2}. 

	%\textcolor{blue}{The exciton polariton array with both zero and $\pi$ phase locking has been realized in a 1D photonic lattice potential at cryostatic temperature \cite{yamamoto}.}
	
	In the present work we realize strong coupling between the excitons of the CsPbBr$_3$ microplates and moir\'e photonic modes in twisted 1D/2D photonic lattice at room temperature, and observe exciton polariton condensation in the moir\'e potential traps. These lattices are composed of two tilted one (two) dimensional lattices fabricated within the microcavity. In our work, the moir\'e exciton polariton zero state and $\pi$ state in these twisted photonic lattices are measured directly. The staggered moir\'e polaritons with different energies are either confined in the potential trap or in the potential barrier region. By modifying the potential trap depth and width of one photonic lattice which results in a moir\'e lattice in the shape of parallelograms, moir\'e polaritons with the distribution of localized dots, dimers and extended lines are observed due to the sensitivity against disorder. Furthermore, we demonstrate and realize the extension of strong coupling between excitons and moir\'e photonic bands in twisted 2D honeycomb lattices. The polariton condensation occurs in the 1D/2D moir\'e potential sites with the coherence time measured. Our work gives clear experimental evidence of phase locked moir\'e polaritons and condensation at room temperature, and paves the way for future investigations into room temperature quantum simulation in moir\'e photonic structures.

The microcavity is fabricated by transferring the chemical vapor deposition (CVD) grown CsPbBr$_3$ microplates onto the bottom DBR. Firstly, we discuss the interaction of excitons with twisted 1D photonic lattices, where the 1D photoresist lattice with the period of 4 $\mu$m (duty cycle: 50:50) and the thickness of 200 nm is formed by using the lithography technique onto the top of perovskite microplates (A layer of SiO$_2$ is deposited onto the perovskite for protection, detail is shown in SM). Another 1D photoresist lattice with the same duty cycle and depth is formed onto the top DBR with the angle of 40$^{\circ}$ against the bottom DBR. The microcavity is fabricated by pasting the top DBR against the bottom DBR with an air gap of around 3 $\mu$m, as shown in Figure 1(a), where the moir\'e stripes consist of periodic rhombus arrays along horizontal ($x$) and vertical ($y$) direction. The optical image of the CsPbBr$_3$ microplate with the twisted moir\'e potential lattice is shown in Figure 1(b). The parameters of the rhombus are shown in the right panel of Figure 1(b), which induces different periodicities along $x$ and $y$ directions.

	In the experiments, we place the sample such that the moir\'e pattern is along the $x$ direction (the long symmetry axis of the rhombus is along the $y$ direction). The energy (\textit{E}) vs wavevector (\textit{$k_x$}) dispersion of the microcavity is measured by using a home-made angle resolved spectroscopy. The pumping source is a femtosecond laser (5700 Hz) with the wavelength of around 400 nm and pulse width of 130 fs. The pumping spot size of the laser is around 40 $\mu$m which covers around 8 periods of the moir\'e lattice. The dispersion along \textit{$k_x$} direction is shown in Figure 1(c). Firstly, we note that there exist several sets of lower branch polaritons (for simplicity, we only plot LP0 and LP1 in Figure 1(c), a larger-energy-range dispersion is plotted in the SM) thanks to the long cavity length where multiple cavity modes strongly couple with the exciton resonance in the perovskite microplates at 2.43 eV. The strong coupling is confirmed at the large wavevector region where the dispersions are clearly bent when approaching the exciton resonance. Other modes with steeper curvatures come from the leakage of photons out of the perovskites or the luminescence from the photoresist. 
	
	The size of the Brillouin zone is reversely proportional to the period of the 1D potential lattice. The dispersion along \textit{$k_y$} direction is shown in Figure 1(d). Due to different periods along \textit{$x$} and \textit{$y$} direction of the moir\'e lattice, the polariton bands show clear different periodicities and amplitudes in Figure 1(c) and Figure 1(d). The change in the size of the first Brillouin zone (from 1.10 $\mu$m$^{-1}$ to 1.29 $\mu$m$^{-1}$) due to different periods along these two directions (from 5.74 $\mu$m to 4.89 $\mu$m) confirms the existence of the coupling between the moir\'e potential traps along the \textit{$x$} direction.

	In addition, in our microcavity, the depth of the moir\'e potential site is around 20 meV, where we observe both the discrete energy levels (the ground state at 2.26 eV, first excited state at 2.27 eV) and the band structure (annotated in Figure 1(c)) along \textit{$k_x$} direction for polariton branch LP1. Within this branch, the discrete energy levels are formed due to small overlap of the lower-energy state between moir\'e potential traps, and polariton bands originate from larger overlap of the higher-energy modes among the moir\'e potential sites. The discrete polariton energy levels and bands are invisible in LP0 due to much smaller photon components. 
	
	Thanks to the coupling between the moir\'e sites, we observe a band gap opening between the ground band and excited band in Figure 1(c) (the ground band structure is invisible due to limited resolution). We fit the band structure of the moir\'e lattices and plot the simulated bands in Figure 2(a) which agree very well with the experimental results. Our simulation also shows the relation between the coupling strength among the moir\'e sites and the band gap opening between the moir\'e ground band and excited band, as discussed in SM. Similar band opening is observed for another lower polariton branch (shown in SM).

	To check the existence of the moir\'e patterns, we measure the energy-resolved polariton distribution in real space. Firstly we show the moir\'e patterns corresponding to the ground state and the first excited states, which are plotted in Figure 2(b-c). In these graphs, we can observe clear moir\'e states with different patterns along \textit{$x$} direction. These modes are confined in the single moir\'e potential traps, where the ground states have a Gaussian shape (the emission at the barrier region comes from the bands from LP2, as shown in the SM), and the first excited modes are two-lobe states with the orientation along \textit{$x$} direction. Due to small overlap between these states trapped in the single moir\'e sites, the dispersions corresponding to these modes are discrete energy levels. The real space image of the modes with a larger energy (2.28 eV) is plotted in Figure 2(d) which corresponds to the moir\'e ground band along {$k_x$} direction. This mode consists of three lobes along \textit{$x$} direction and promises larger coupling between the nearest potential traps. With higher energy, moir\'e excited bands are observed along \textit{$k_x$} direction in Figure 1(c) and Figure 2(a).

	\begin{figure}
		\centering
		\includegraphics[width=0.95\linewidth]{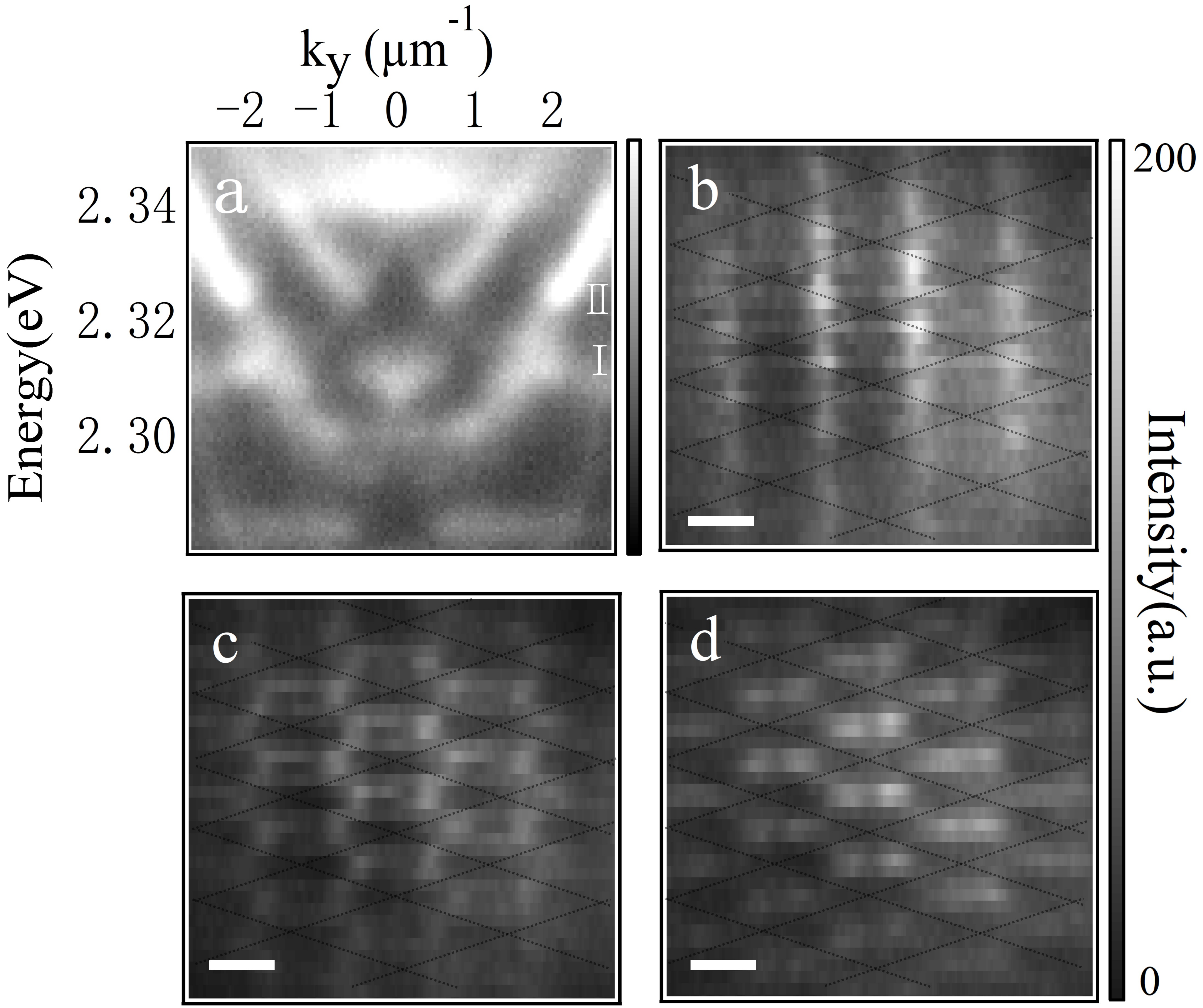}
		\caption{\textbf{Phase locked moir\'e polariton patterns.} (a) Zoomed dispersions along along $k_y$ direction. Band I and II are labeled. (b-d) Moir\'e polariton spatial distributions at the bottom of band 1, the top of band I and the bottom of band II shown in Figure 3(a).  The dash lines represent two photonic lattices. Scale bars in (b, d): 5 $\mu$m.}
	\end{figure}
	
	\begin{figure}
		\centering
		\includegraphics[width=\linewidth]{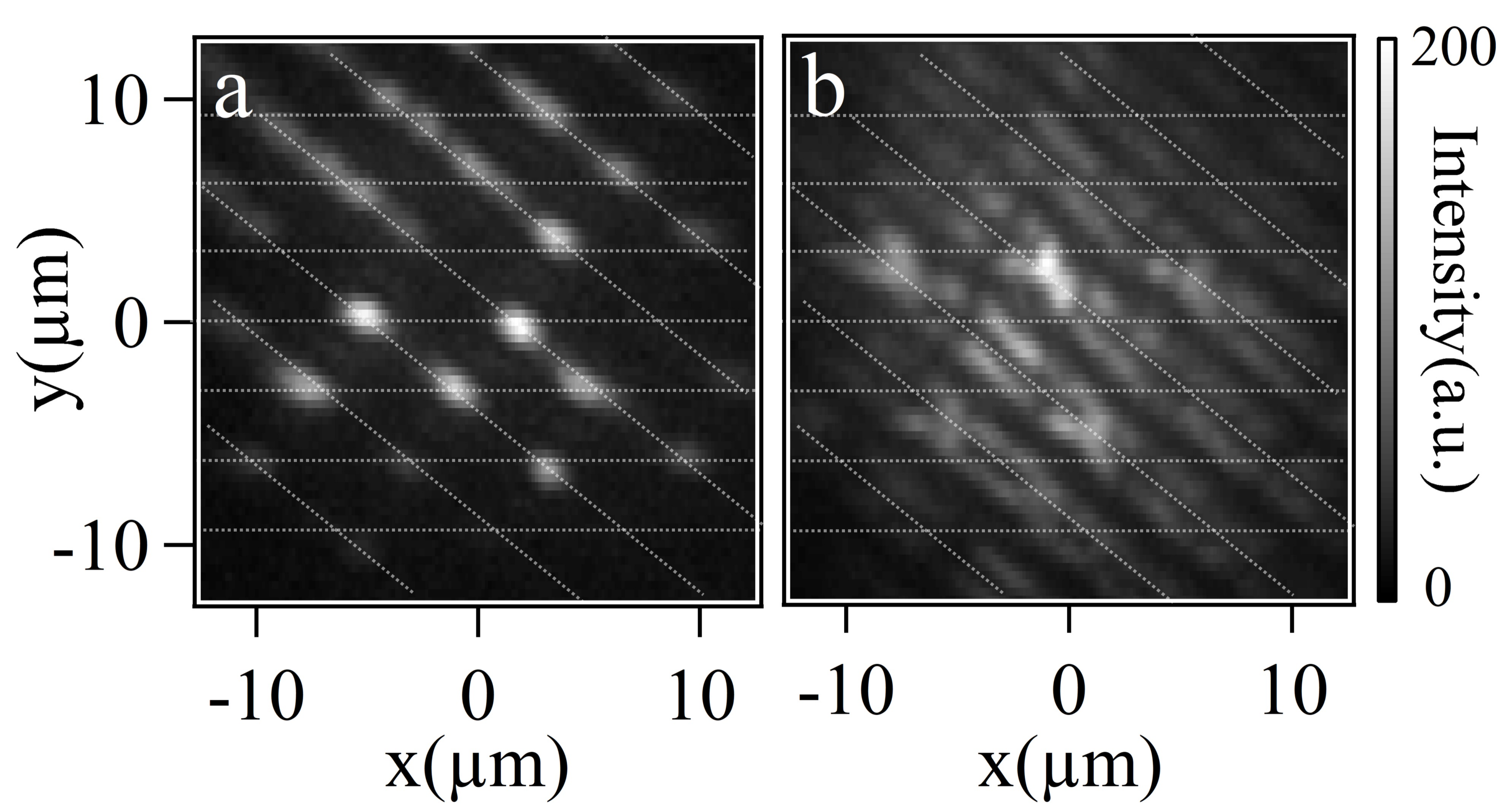}
		\caption{\textbf{Moir\'e polariton patterns in a modulated 1D and twisted 2D photonic lattice.} (a) Schematics of the moir\'e pattern in modulated 1D lattices. (b) The zoomed image of the modulated 1D twisted photonic lattice. (c, d) The spatial distribution of the moir\'e polariton ground state and excited state. (e) Schematics of the moir\'e pattern in twisted 2D lattices. (f-h) The spatial distribution of the moir\'e polariton ground state, and higher excited states, with the dispersion shown in SM. Scale bars in (b-d, f-g): 5 $\mu$m.} 
	\end{figure}
	
	\begin{figure*}
		\centering
		\includegraphics[width=0.8\linewidth]{./Fig5.pdf}
		\caption{\textbf{Moir\'e polariton condensation in twisted 1D/2D photonic lattices.} (a, e) Integrated intensity and linewidth of the moir\'e polaritons against the pumping density in twisted 1D(a)/2D(e) lattices. (b, c, f, g) Dispersion and real space images of the moir\'e polariton condensate in 1D(b, c)/2D(f, g) lattices, scale bars in (c, g): 5 $\mu$m. (d, h) The first-order temporal coherence $g^{(1)}$($\tau$) of 1D(d) and 2D(h) photonic lattice. }
	\end{figure*}
	
	On the other hand, the band gap opening is also clearly observed along \textit{$k_y$} direction, as shown in Figure 1(d) and Figure 3(a). The moir\'e polariton modes at the bottom (top) state of the band I and the bottom state of the band II shown in Figure 3(a) are plotted in Figure 3(b, c, d) (the two bands correspond to the ground and excited band along \textit{$k_y$} direction). From these, we find clear zero and $\pi$ phase locking of these high-energy modes across the moir\'e polariton lattice. Note that the zero state appears when the phase difference of polaritons in the moir\'e lattice is zero, which will give nonzero intensity distribution with the wavevector of $k_{y}=\pm m\pi /a$ (where $m$ is an even integer). On the other hand, the $\pi$ state can be observed with the wavevector of $k_{y}=\pm n\pi /a$ (where $n$ is an odd integer) \cite{yamamoto}. At the energy of 2.30 eV which corresponds to the zero state located at the bottom of band I with the wavevector of 0 and $\pm 1.1\ \mu$m$^{-1}$, Figure 3(b) shows a clear moir\'e pattern along the \textit{$y$} direction. In Figure 3(c) and (d) we can observe different moir\'e patterns at the top of the band I and the bottom of band II (the energy of these two states are 2.31 eV and 2.32 eV with the wavevector of $\pm 0.55\ \mu$m$^{-1}$ and $\pm 1.65\ \mu$m$^{-1}$), which is the $\pi$ state. In above moir\'e zero state, the polaritons are visible in both the potential trap and barrier of the moir\'e lattice due to finite overlap of the wavefunctions, as plotted in Figure 3(b). Whereas the moir\'e $\pi$ states shown in Figure 3(c) and Figure 3(d) are composed of a higher energy mode showing a staggered pattern. More importantly, these states are mainly located at the barrier region of the moir\'e lattice with the same periodicity as the low-energy modes in Figure 2(b-d).

Above moir\'e polaritons are robust against the disorders. In SM we show the spatial modes of the moir\'e ground state, excited state and moir\'e bands at other positions. Although distorted partly due to the inhomogeneity within the microcavity, they still keep the structure inherited from the moir\'e lattice, the phase locked polariton distributions in the moir\'e bands are similar as Figure 2 and Figure 3.

	The moir\'e polaritons can be engineered quite easily within our photonic structures. In the twisted 1D lattices, the energy depth and width of the moir\'e potential sites can be tuned by modifying the parameters of one photonic lattice. In this case, different moir\'e polariton states can be observed directly. We replace the photoresist lattice on the perovskite microplate with a PMMA layer (thickness: 230 nm) and fabricate the photonic lattice (period: 4 $\mu$m, potential trap: 1.6 $\mu$m, potential barrier: 2.4 $\mu$m, depth: 50 nm) by using laser direct writing technique. In the new structures, the moir\'e polariton ground state distribution is in the shape of a parallelogram (Figure 4(a, b)). Due to lower potential trap depth of the PMMA photonic lattice, these moir\'e polaritons are fragile against defects or disorder due to lower energy and become deformed with random distribution, for example, localization trapped in the moir\'e potential, dimers, or delocalized lines, as shown in Figure 4(c). Excited and free moir\'e polariton modes in Figure 4(d) are robust against disorder in the shape of delocalized lines along the direction of the top photonic lattice (Polariton spatial modes at another position is shown in SM). In addition, our experimental scheme to observe moir\'e polaritons can be extended to twisted 2D lattices. To realize this, we fabricate a 2D honeycomb lattice onto the top of the perovskite microplates and the top DBR. With the twisting angle of 11$^{\circ}$ as shown in Figure 4(e), we can observe different moir\'e polariton patterns as shown in Figure 4(f-h) (the dispersions plotted in the SM). The observation of these tunable moir\'e polaritons enables studies of novel polariton phase transitions at room temperature in future works.

	Finally, exciton polariton condensation can be realized in above 1D/2D twisted photonic lattices by increasing the pumping density. In Figure 5(a, e) we plot the integrated intensity and linewidth of polariton modes against the pumping density of the 1D/2D twisted lattices, which show polariton condensation occurs at around 11.5 $\mu$J/$cm^2$ (1D) and 12 $\mu$J/$cm^2$ (2D). In the two lattices, polaritons condense at the ground state of LP0 (Figure 5(b, f)) and are confined at the moir\'e potential traps in the 1D/2D photonic lattices, which are shown in Figure 5(c, g). We find the coherence time of the polaritons condensate in the two twisted lattices is around 1.4 ps, which is obtained by measuring the first-order temporal coherence $g^{(1)}$($\tau$)\cite{chen apl}, plotted in Figure 5(d, h).

Compared with exciton polaritons loaded into periodic photonic lattices by etching pillars or growing buried mesa arrays \cite{polariton lieb1, polariton lieb2} in GaAs microcavity and moir\'e exciton polaritons based on TMD monolayer heterostructures at low temperature \cite{moire polariton}, we can directly measure the moir\'e polariton ground state, moir\'e excited state, moir\'e polariton bands and moir\'e polariton condensation at room temperature. This allows to observe how the moir\'e polaritons are manipulated by the periodicity, the potential depth and other parameters much more easily.

	To summarize, we observe strong coupling between excitons of CsPbBr$_3$ microplates and moir\'e photonic bands of 1D (2D) twisted lattice in a microcavity at room temperature. Different moir\'e polariton stripes corresponding to the discrete energy levels and the continuous bands are observed. Moir\'e polariton condensation in the 1D/2D lattices is shown and coherence time is measured. The realization of moir\'e polariton arrays with different phase locking can be used to investigate quantum phase transitions between the superfluid and Mott insulators in twisted photonic lattices at room temperature in the future.

	%\section*{Acknowledgments}
	\begin{acknowledgments}
		T.Gao acknowledges the support from the National Natural Science Foundation of China (NSFC, No. $12174285$, $12474315$), especially T.Gao wants to thank Yongyou Zhang and Yang Yang for fruitful discussion. The Paderborn group acknowledges support from the German Research Foundation (DFG, No. 519608013). H.Dai thanks the National Natural Science Foundation of China (NSFC, No. $62375200$).
	\end{acknowledgments}

	%plain tex reference style

\end{document}